\documentclass{article}

\usepackage[final]{neurips_2023}

\usepackage[utf8]{inputenc} 
\usepackage[T1]{fontenc}    
\usepackage{hyperref}       
\usepackage{url}            
\usepackage{booktabs}       
\usepackage{amsfonts}       
\usepackage{nicefrac}       
\usepackage{microtype}      
\usepackage{xcolor}         
\usepackage{subfig}

\author{%
  Humanity Unleashed
}

\usepackage[english]{babel}

\usepackage{soul}
\usepackage{natbib}
\usepackage{amsmath}
\usepackage{graphicx}
\usepackage{color}
\usepackage{seqsplit}
\usepackage{xargs}
\usepackage{xstring}
\usepackage{commands}

\usepackage{xcolor} 
\newcount\Comments  

\title{Creating a Cooperative AI Policymaking Platform through Open-Source Collaboration} 
\usepackage{microtype}
\usepackage{graphicx}
\usepackage{booktabs} 

\newcount\Comments  
\newcommand{\kibitz}[2]{\ifnum\Comments=1\textcolor{#1}{#2}\fi}
\newcommand{\ez}[1] {\kibitz{orange} {[EZ: #1]}}

\definecolor{ultramarine}{rgb}{0.07, 0.04, 0.56}

\definecolor{kindaultraviolet}{rgb}{0.6, 0.29, 0.54} 
\newcommand{\hg}[1] {\kibitz{kindaultraviolet} {[HG: #1]}}

\definecolor{randomnicelookingcolor}{rgb}{0.7, 0.8, 0.9}

\definecolor{darkgreen}{rgb}{0.0, 0.5, 0.0}

\definecolor{darkblue}{rgb}{0.0, 0.0, 0.5}

\definecolor{jlcolor}{rgb}{0.5, 0.5, 0.0}

\usepackage{hyperref}
\usepackage[]{mdframed}
\usepackage[subtle]{savetrees}

\usepackage{amsmath,amsfonts}

\def\eqref#1{equation~\ref{#1}}

\def\1{\boldsymbol{1}}

\DeclareMathAlphabet{\mathsfit}{\encodingdefault}{\sfdefault}{m}{sl}
\SetMathAlphabet{\mathsfit}{bold}{\encodingdefault}{\sfdefault}{bx}{n}

\newcommand{\states}{S}

\newcommand{\action}[1][]{\ifthenelse{\equal{#1}{}}{\boldsymbol{a}}{a_{#1}}}
\newcommand{\actiont}[1][]{\ifthenelse{\equal{#1}{}}{{\boldsymbol{a}}_t}{a_{#1,t}}}

\newcommand{\type}[1][]{\ifthenelse{\equal{#1}{}}{\theta}{\theta_{#1}}}
\newcommand{\typet}[1][]{\ifthenelse{\equal{#1}{}}{{\boldsymbol \theta}_t}{\theta_{#1,t}}}
\newcommand{\types}[1][]{\ifthenelse{\equal{#1}{}}{\Theta}{\Theta_{#1}}}

\begin{document}

\maketitle

\vspace{15pt}
\begin{abstract}
Advances in artificial intelligence (AI) present significant risks and opportunities, requiring improved governance to mitigate societal harms and promote equitable benefits. Current incentive structures and regulatory delays may hinder responsible AI development and deployment, particularly in light of the transformative potential of large language models (LLMs). To address these challenges, we propose developing the following three contributions: (1) a large multimodal text and economic-timeseries foundation model that integrates economic and natural language policy data for enhanced forecasting and decision-making, (2) algorithmic mechanisms for eliciting diverse and representative perspectives, enabling the creation of data-driven public policy recommendations, and (3) an AI-driven web platform for supporting transparent, inclusive, and data-driven policymaking.

\end{abstract}
\section{Introduction}



As advances in artificial intelligence continue to reshape our world, humanity must grapple with new forms of risk and opportunity. Concerns abound regarding rogue AI—whether arising from malicious human actors or from autonomous agents themselves \citep{9245153,AI_Safety_2024} —as well as the transformative power that AI confers on corporate and state competitors \citep{khan2019maintaining,harraf2021potentials,haner2019artificial,morris2024bigtech}. While technical work on alignment and safety is essential, it only succeeds insofar as we can ensure that those who control AI systems use them responsibly. Yet our current incentive structures — shaped heavily by short-term profit motives and often insufficiently guided by longer-term societal well-being \citep{szczepanski2019economic, acemoglu2022tasks,  zajko2022artificial, dhar2020carbon, wang2024ecological} — hinder effective governance. Regulation often lags behind until an immediate negative event occurs. In light of the recent advancements in scaling LLMs, regulatory lag poses considerable societal risk \citep{hendrycks2023overview, turchin2020classification}.



Improving our institutional capacities-both for thoughtful regulation and for productive application of advanced AI—is crucial \citep{zhang24sed}. 
Rapid, data-driven policy interventions are needed not only to mitigate existential risks but also to ensure that the coming shift to an AI-driven economy yields broadly shared benefits.

We believe one promising avenue is to harness and repurpose emerging AI methods—especially “foundation models”—to assist policymakers directly. As AI grows more capable, open-source models aligned with public welfare can help guide regulation and inform evidence-based policy decisions. By developing these tools transparently and collaboratively, we can avoid profit-driven biases and build trust with both policymakers and the public \citep{osterloh2004trust}. We have thus taken this approach, believing that progress made in the private sector can be turned towards public benefit, enabling timely, data-driven policies that safeguard vulnerable populations and the commons through the open-source and academic community.

 A critical component of this vision is the development of a specialized “time-series LLM.” In our context, this refers to a large language model that integrates macroeconomic and microeconomic time-series data—such as GDP, inflation rates, and legislative histories—with textual information like policy documents and news. Such a model can forecast economic indicators, predict policy impacts, and support scenario analysis with enhanced accuracy. By combining textual reasoning with temporal, quantitative forecasting \citep{liu2024time, williams2024context}, we can obtain a richer understanding of how policy choices shape economic and social outcomes. This capability will be paired with a “value elicitation” mechanism to discern public preferences and an LLM-based public policy generator to propose policies aligned with these values. 
 
 Together, these components can strengthen the policymaking process and open new approaches toward effective, inclusive governance. We have structured ourselves as an open-source effort in pursuit of these capabilities and encourage all motivated or interested in navigating these challenges for collective benefit to participate; we warmly welcome additional contributors over the coming months.


\section{Project Overview}

Our initiative, led by Humanity Unleashed, a 501(c)3 nonprofit, unites numerous independent researchers - academics, students, and nonprofit staff - into a large-scale, open-source collaboration. By organizing the proposed research into multiple subprojects with more conventionally sized teams and supporting them through a nonprofit framework, we overcome the traditional coordination challenges faced by academia. This approach fosters rapid, iterative progress and supports the eventual integration of all subcomponents into a functional prototype — a cooperative AI policymaking platform capable of garnering public trust and support.

The integrated platform will feature the following components:

\textbf{\hyperref[sec:econ_transformer]{Economics Transformer}}: A multimodal time-series language model trained on macroeconomic and microeconomic data, coupled with legislative and event information, capable of forecasting economic outcomes and quantifying uncertainty. This goes beyond classical models (e.g., DSGE \citep{christiano2018dsge}) and provides richer, data-driven insights on policy impacts.

\textbf{\hyperref[sec:ai_legislator]{AI Legislator}}: A mechanism that first elicits user values via a carefully designed questionnaire and hierarchical Bayesian modeling and then generates policy proposals that reflect these values. This ensures output policies have broad, bipartisan appeal while considering nuanced public preferences.

\textbf{\hyperref[sec:policy_interface]{Policy Interface}}: A unified, user-friendly interface that enables policymakers, researchers, and the public to interact with these tools. Through open-source release of code, datasets, and benchmarks, we invite the broader community to refine, extend, and evaluate these models.

By leveraging a nonprofit setting and academic collaboration, we align incentives toward public benefit rather than profit. Our 2024-2025 roadmap includes completing team organization, data gathering, model development, and iterative refinements, culminating in machine learning conference submissions in mid 2025 and the public release of our integrated tool in late 2025.

In summary, we aim to advance the state of AI-assisted policymaking. Our contributions will establish open-source baselines, benchmarks, and a principled method for integrating time-series economic forecasting with LLM-based value elicitation and policy generation. Ultimately, we seek to demonstrate a constructive, transparent, and inclusive approach to using AI in support of governance, social welfare, and long-term societal resilience.

\begin{figure}
    \centering
    \includegraphics[width=1\linewidth]{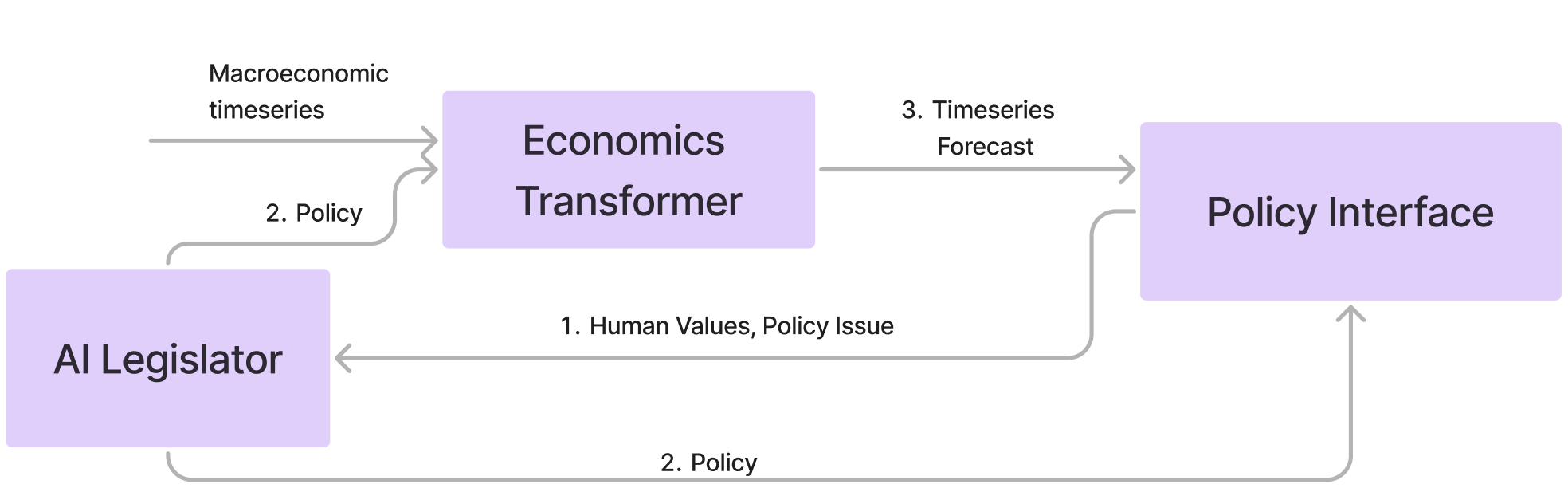}
    \caption{A broad overview of the system’s workflow. The AI Legislator first collects user values and policy objectives from user interactions mediated by the policy interface (arrow 1) and transmits them to both the Policy Interface and the Economics Transformer (arrow 2). The Economics Transformer also receives macroeconomic time-series data from the Federal Reserve, and uses these inputs to generate forecasts, informing the effect of policies on various macroeconomic indicators like GDP over time. These forecasts, along with the proposed policies, are then routed back to the Policy Interface (arrow 3), where policymakers, researchers, and the public can interact with, refine, and better understand policy.}
    \label{fig:main}
\end{figure}

For the remainder of the report, we provide an in-depth overview of each component pictured in Figure~\ref{fig:main}: the Economics Transformer (Section~\ref{sec:econ_transformer}), the AI Legislator (Section~\ref{sec:ai_legislator}), and the Policy Interface (Section~\ref{sec:policy_interface}). We conclude with discussing a subproject which will study the societal impact of LLMs within the United States economy in Section~\ref{sec:societal_impact}, concerning key implications of our project initiative.

\section{Economics Transformer: Enhancing LLMs for Predicting Policy Impacts on Multivariate Time Series}
\label{sec:econ_transformer}
The Economics Transformer is designed to predict multivariate time series \textit{in the context of a proposed policy or event} given in the form of natural language. Central to our approach is leveraging existing LLMs by fine-tuning them to predict multivariate time series while retaining their language understanding capabilities. This project is divided into several subprojects—data collection, model architecture, scaling laws, and evaluation frameworks—to tackle the distinct challenges involved in training the Economics Transformer. We motivate each of these subprojects below.

\textbf{Data Collection.} To support the development of the Economics Transformer, we will curate a dataset comprising paired \textit{(policy, time series)} examples that reflect the dynamics of economic variables in response to specific policies or events. Although numerical economic time-series and other relevant textual data (eg. economic news, policy announcements, reports, etc.) are readily available from public sources, it is less clear how to effectively integrate data across these two modalities and establishing meaningful pairings that reinforce a strong causal connection between the natural language context and time series data. More details about our approach for data collection is given in Section~\ref{subsec:econ_transformer_data_collection}.

\textbf{Model Architecture.} Approaches such as those in \citep{requeima2024llm, williams2024context} have demonstrated that leveraging pretrained LLMs in \textit{zero-shot} settings with time series and text can produce reasonable results for context-driven forecasting. Although this is a reasonable baseline, developing novel architectures can introduce a beneficial inductive bias for our downstream use case. Designing such an architecture includes joint encoders to manage the multimodal and multivariate nature of the input, along with components which ensure the Economics Transformer produces output distributions quantifying the uncertainty of its predictions.
In Section~\ref{subsec:econ_transformer_model_architecture} we outline our proposed approach for the general underlying model architecture.

\textbf{Scaling Laws.} In line with prior work on fitting neural scaling laws to model performance based on parameters, compute, and dataset size~\citep{kaplan2020scaling, edwards2024scaling} we aim to study how performance in multimodal time-series and language settings is influenced by these factors. Specifically, we aim to investigate the potentially distinct effects of the two modalities and assess how dataset quality influences downstream performance. Further details on our considerations for the scaling laws of multimodal time-series language models and our planned investigation into hyperparameter transfer approaches are provided in Section~\ref{subsec:econ_transformer_scaling}.

\textbf{Evaluation Frameworks.}
For evaluation, we will focus on benchmark datasets that reflect real-world macroeconomic forecasting challenges. Instead of relying on canonical datasets that may not be policy-relevant (e.g., generic traffic or weather datasets), we will concentrate on macroeconomic and finance-related benchmarks such as exchange rate forecasting \citep{godahewa2021monash} and hold-out sets from the Federal Reserve Bank of St. Louis Time Series (FRED) dataset \citep{fred2011dataset}. Recognizing potential concerns about pretrained LLMs' prior exposure to these datasets as well as the need for unified frameworks to evaluate language model capabilities~\citep{eval-harness, lambert2024rewardbench}, we similarly plan on releasing a live benchmarking leaderboard which stays up-to-date with the latest FRED-MD dataset and allows for performance comparisons across traditional methods and novel approaches. More details are given in Section~\ref{subsec:econ_transformer_evaluation}.



\subsection{Data Collection for the Economic Transformer}
\label{subsec:econ_transformer_data_collection}

This project aims to create a comprehensive, multimodal dataset encompassing numerical economic time-series and long-form textual policy data. By addressing the limitations of existing datasets, this work will enable innovative insights into policy evaluation and economic forecasting. 

Economic research and policymaking rely on robust datasets to analyze trends, simulate interventions, and evaluate outcomes. However, we identify significant limitations within existing datasets: 
\begin{itemize}
    \item \textbf{Fragmentation:} Numerical and textual data are often silo-ed, making it difficult to perform cross-modal analyses.
    \item \textbf{Limited Coverage:} Many datasets lack generalization, instead focusing on specific regions, industries, or time periods. 
    \item \textbf{Lack of Contextualization:} Textual narratives and long-form policy documents, which provide critical context, are rarely integrated with numerical time-series data
\end{itemize}

Our proposed dataset will fill these gaps by collecting, cleaning, and harmonizing economic time-series data across modalities, ensuring greater accessibility and usability. The significance of this work lies in its ability to bridge the gap between fragmented economic datasets and integrate them with textual policy datasets. This requires the dataset to have the following two key components.

\textbf{Numerical Economic Time-Series Data:} This component will aggregate structured numerical data from publicly available sources such as central banks, government agencies, and international organizations. Examples include GDP, unemployment rates, inflation indices, trade balances, and market indicators. 
\begin{itemize}
    \item \textbf{Granularity:} Inclusion of data at multiple levels (e.g., country, regional, and sectoral levels).  
    \item \textbf{Historical Coverage:} A focus on extending time horizons to provide historical context for long-term analysis.  
\end{itemize}

\textbf{Long-Form Economic/Policy Textual Data: } This component will focus on in-depth policy documents, white papers, and research reports.
\begin{itemize}
    \item \textbf{Topic Classification:} Tagging documents with economic themes, such as taxation, monetary policy, or trade agreements.   
    \item \textbf{Historical Coverage:} Emphasis on the inclusion of diverse policies from all political administrations.
\end{itemize}

The proposed dataset will be constructed through the use of established public repositories such as the Federal Reserve Economic Data (FRED), and the GDELT Event Database.

To allow for the forecasting of economic time-series given a natural language policy, our approach will temporally align textual and time series data, synchronizing policy events with corresponding economic metrics based on timestamps or event markers.

\textbf{Temporal Alignment}: Standardize timestamps across numerical and textual data to enable temporal pairing across modalities. This involves mapping each piece of text \( x \) and time series segment \( y \) to a common time index \( t \): $\mathcal{A}: (x_t, y_t) \quad \forall t \in T$, where \( T \) is the set of all time indices \citep{baltrusaitis_multimodal_2019}.

\textbf{Event-Based Alignment}: Align data around significant events, such as policy enactments, using event indicator functions:
    \begin{equation}
    E(t) = \begin{cases}
    1, & \text{if an event occurs at time } t \\
    0, & \text{otherwise}.
    \end{cases}
    \end{equation}

This allows the model to focus on the periods where interactions between modalities are most critical \citep{poria_multimodal_2018}.

\subsection{Novel Architecture Explorations: Foundation Model for Multi-Modal Time Series Tasks}
\label{subsec:econ_transformer_model_architecture}

\paragraph{Project Summary}

This research project aims to develop an innovative foundation model for practical time series analysis, including forecasting~\cite{cao2020spectral, cao2021spectral, cao2023large, niu2022mu2rest, niu2023time}, anomaly detection~\citep{zhao2020multivariate}, causal inference~\citep{zhang2022counterfactual, cao2023estimating, pmlr-v235-huang24r} and physics informed generation~\citep{meng2022physics, griesemer2024active, xiao2023coupled}, addressing critical limitations in current approaches. Towards this end, our proposal has two key directions: firstly, we will develop a novel fine-tuning approach that integrates multivariate time series analysis capabilities into existing large language models, effectively harnessing their innate ability to process and understand textual data. Specifically, we propose to modify Time-LLM \citep{jin2024timellmtimeseriesforecasting}, a model designed to handle time series with textual context, to perform multimodal, multivariate probabilistic forecasting. Secondly, we will propose Continuous-Valued Transformer (CVT) model which extends recent advancements in transformer architectures and diffusion models to handle continuous-valued time series data alongside textual information. By incorporating a novel Diffusion Loss 
function and adaptive temporal resolution mechanisms, the CVT model promises to capture complex temporal dependencies across various scales and modalities. The significance of this work lies in its potential to revolutionize time series analysis across multiple domains, enhancing decision-making processes in AI-driven policymaking platforms and improving forecasting accuracy in critical sectors.



\paragraph{Background and Objectives}
The primary question this project addresses is: How can we develop a foundation model for multivariate time series that effectively integrates with large pre-trained models while optimizing for accuracy, multimodal capabilities, uncertainty quantification, and counterfactual generation? This question is crucial due to the ubiquity of time series data across various domains and the increasing need for models that can handle complex, multimodal data~\citep{jia2024gpt4mts}. In fields such as finance, healthcare, and climate science, the ability to accurately predict and interpret temporal patterns while leveraging textual information can lead to groundbreaking insights and improved decision-making. The question is both interesting and challenging due to several factors: aligning different data modalities without losing their specific characteristics, developing a tokenization method that preserves both global and local temporal information, capturing complex inter-temporal and cross-feature dependencies while maintaining computational efficiency, and balancing discrete time series representation with continuous modeling. 


Specifically, time series foundation models face significant challenges in effectively representing complex, multi-dimensional temporal data while preserving both local and global information~\citep{das2023decoder, woo2024unified,talukder2024totem,yue2022ts2vec, ansari2024chronos,rasul2024lagllama}. 
The integration of time series analysis with textual information, inspired by the success of large language models (LLMs) such as LLaMA~\citep{touvron2023llama} and ChatGPT-4~\citep{openai2023gpt}, presents additional technical challenges. Current approaches struggle to effectively process and combine these disparate data types while maintaining their unique characteristics. There is a need for novel multimodal architectures that can seamlessly handle both numerical time series and associated textual data. Such architectures must address several key technical requirements: 
(1) Developing a unified representation that preserves the temporal structure of time series and the semantic content of text, (2) Designing attention mechanisms that can capture dependencies both within and across modalities, (3) Creating training objectives that balance the learning of temporal patterns and textual understanding, (4) Ensuring the model's ability to generalize across different time scales and data distributions.
Addressing these technical challenges is essential for creating a versatile foundation model capable of advanced time series analysis in conjunction with textual data processing.

\paragraph{Research Direction 1: Fine-Tuning LLMs for Multimodal Ability}
\label{subsec:econ_transformer_multivariate}
Time-LLM has demonstrated the effectiveness of reprogramming large language models for time series forecasting, outperforming traditional time series models in certain univariate forecasting tasks \citep{jin2024timellmtimeseriesforecasting}. However, Time-LLM processes each variate dimension individually during patch embedding, leading to a loss in cross-variate dependencies crucial for multivariate forecasting. Our research addresses this limitation by proposing two distinct approaches to modifying Time-LLM:

    \textbf{I. Incorporating Cross-Variate Information in Patch Embedding:}
    Similar to UniTST \citep{liu2024unitsteffectivelymodelinginterseries}, this approach combines patches into a 2D tensor that spans both temporal and variate dimensions. Linear projection with positional embeddings is applied on the tensor to generate 2D token embeddings, which are then flattened into a 1D token matrix compatible with Time-LLM's reprogramming layer. 
    
    \textbf{II. Capturing Cross-Variate Information During Reprogramming:}
    Inspired by Crossformers \citep{zhang2023crossformer} and CARD \citep{xue2024cardchannelalignedrobust}, this approach replaces the cross-attention mechanism in the reprogramming layer with two-stage attention, which individually captures and accumulates temporal and cross-variate dependencies without modifying the patches. 

\paragraph{Research Direction 2: Designing Time Series Multimodal Model} Our research aims to revolutionize the field of time series analysis by developing a state-of-the-art foundation model that addresses current limitations while leveraging the power of transformers and integrating textual information. The proposed approach builds upon recent advancements such as TEMPO~\citep{cao2024tempo}, TimeDiT~\citep{cao2024timedit} and MAR~\citep{li2024autoregressive}, extending their capabilities to create a more versatile and powerful model for time series processing.
In this proposal, we propse the Continuous-Valued Transformer (CVT) to operate directly on the continuous time series domain. This approach overcomes the limitations of discrete tokenization methods, which have been a significant bottleneck in applying transformer architectures to time series data. The CVT employs a continuous embedding layer that maps input time series values to a high-dimensional continuous space, preserving the richness and nuance of the original data while enabling transformer-style processing. A key component of our model is the "Diffusion Loss" function, which replaces the traditional categorical cross-entropy loss used in discrete approaches. This loss function draws inspiration from diffusion models but is tailored for efficient training within the transformer framework. It models continuous probability distributions for each time step using a mixture of Gaussians, allowing for multi-modal predictions that capture the inherent uncertainty in probabilistic time series forecasting. 
To handle the complexities of multivariate time series, we implement an adaptive temporal resolution mechanism that allows the model to automatically adjust its focus on different time scales within the input sequence. This is particularly crucial for dealing with time series data that exhibit patterns at various frequencies, from high-frequency trading data to slow-moving economic indicators. 
Our multi-modal CVT model architecture features modality-specific encoders for time series and text, based on our enhanced TEMPO model and LLaMA (or similar large language models) respectively. The fusion encoder shares weights across modalities, further encouraging alignment in the shared embedding space. 

Specifically, to effectively integrate language and time series data, we turn towards joint encoding techniques that map both modalities into a shared latent space; specifically, multimodal autoencoders with modality-specific encoders to transform language data \( x \) and time series data \( y \) into unified representations: $z = f_{\text{enc}}^{\text{lang}}(x) = f_{\text{enc}}^{\text{ts}}(y).$ In this setup, \( f_{\text{enc}}^{\text{lang}} \) and \( f_{\text{enc}}^{\text{ts}} \) are encoders for the language and time series data, respectively, both producing the same latent representation \( z \in \mathbb{R}^{d_z}\) \citep{10.5555/3104482.3104569}. This shared latent space captures common features and correlations between the modalities, enabling the model to learn intricate relationships between policy texts and economic indicators.
To analyze the relationship of each modality's affect on the other, we will employ a cross-modal attention mechanism to allow the model to dynamically focus on relevant features across both modalities. Within the transformer architecture, we compute attention scores between modality-specific queries, keys, and values.

\subsection{Scaling Laws for Multimodal Time-Series Language Models} 
\label{subsec:econ_transformer_scaling}
Current work often treats time series data in isolation, leaving gaps in understanding how different modalities (like natural language and time series) impact model performance when integrated \citep{liu2024timemmd}. The scaling goals are twofold: first, to establish how increasing data, compute, and model size influences performance in multimodal settings; and second, to examine how the proportion of different modalities within a dataset affects forecasting accuracy and optimal hyperparameters. Previous research \citep{aghajanyan23a} indicates that evaluation loss in multimodal models is influenced by the complexity of combining different datasets and the additional computational demands this entails. This relationship for two modalities can be expressed as $$\mathcal{L}(N, D_i, D_j) = \left[ \frac{\mathcal{L}(N, D_i) + \mathcal{L}(N, D_j)}{2} \right] - \mathcal{C}_{i,j} + \frac{A_{i,j}}{N^{\alpha_{i,j}}} + \frac{B_{i,j}}{|D_i| + |D_j|^{\beta_{i,j}}},$$ where $\mathcal{L}(N, D_i, D_j)$ is the evaluation loss for model parameters $N$ and datasets $D_i$ and $D_j$, $\mathcal{C}_{i,j}$ represents the information gain from combining datasets, $A_{i,j}$ and $B_{i,j}$ are constants related to model complexity and data, and $\alpha_{i,j}$ and $\beta_{i,j}$ are scaling exponents \citep{aghajanyan23a}. For time series models, the evaluation loss often \hg{I don't like "often" here -- not scientific language.}follows a power-law scaling relationship, $\mathcal{L}(A) = \left( \frac{A}{A_0} \right)^{-B_0}$, where $A$ is the scaling factor (such as data size, compute, or model size), and $A_0$ and $B_0$ are normalization and fitting constants \citep{edwards2024scaling}. To determine how these two metrics will interact, this project will use a transformer model \hg{not sure this is the right citation here lol}\hg{it now says vaswani 2023 so date is a tad off} \ez{they've updated the vaswani paper multiple times, 2023 is fine} \citep{vaswani2023attention} and various multimodal time series datasets like TIME-MMD \citep{liu2024timemmd} and FRED \citep{fred2011dataset}. Initial steps involve encoding time series data for use in the transformer, experimenting with encoding methods like vector quantization \citep{rasul2024vqtr} and a direct mapping from the continuous time series data to a transformer’s continuous representation of space\hg{stuff like " a direct mapping from the continuous time series data to a transformer’s continuous representation of space" is imo way too low-density. Do you just mean "point-wise tokenization"?}, and then determining the optimal relationship between compute, data, and model size when using time series data to minimize prediction errors. The project will also investigate the optimal composition of modalities within datasets and fine-tune hyperparameters using approaches like $\pi$-tuning \citep{WuWGLZSL23} and $\mu$-P \hg{let's use greek letter mu}transfer \citep{yang2021tuning}. By optimizing resource allocation across data amount and type, number of model parameters, and compute; this project aims to achieve more optimal multimodal prediction performance  with fewer resources, making prediction feasible for larger domains and datasets through saving much in compute resources \hg{"hundreds of thousands" is a figure pulled out of nowhere. Is this accurate? If so, what calculation led to this?} \ez{i rm'd the hundred of thousands}.


\subsection{DBITS: Dynamic Benchmarking of Indicator Time Series}\label{subsec:econ_transformer_evaluation}

\textbf{Project Description.} To inform policymakers to make better policies, we need better forecasts of economic indicators (e.g., GDP). A proposed solution for this is a multimodal time series forecasting transformer, but there is no way to dynamically evaluate this compared to more traditional models in real-time. This proposal aims to create a live benchmarking leaderboard for economic time series models. Having a live leaderboard with different evaluation filters and a history of relative performance to point out trends will help researchers develop better models for predicting economic indicators where changes in culture and politics can make a significant impact on trends in the data as better time series models for economic indicators can be made through ensemble methods such as bagging and boosting from having a reliable ranking of models in different contexts. \ez{a lot of this info is redundant from prior sections}

This continually updating live leaderboard will be realized by creating a new open-source hosted code repository that will automatically scrape the latest FRED-MD dataset in real time ~\citep{mccracken_fred-md_2016}, run the appropriate evaluations (evals) for all the models, and display their rankings on a continually updating basis with the option to select different evaluation filters. This solution will also have the ability to upload a model via uploading a script that the server can use to run evals. 

\textbf{Background and Technical Need}
 The current state-of-the-art for time series leaderboards is the OpenTS leaderboard \citep{qiu2024tfb} which benchmarks various time series models and serves as a great foundation for our own new live leaderboard as it also incorporates a static snapshot of the FRED-MD dataset. However, it does not update in real-time automatically \citep{qiu2024tfb}. For the OpenTS leaderboard there is currently significant overhead for people looking to compare the performance of different time series models as they would have to setup their own copy of the leaderboard to get live results and/or add their own models \citep{qiu2024tfb}. This requires both time and expertise in terms of software development as well as the necessary supporting infrastructure to host and deploy a local version of the backend. Another recent leaderboard is GIFT from Salesforce AI where they allow users to select different horizons for the ranking \citep{aksu2024gifteval}. However, they too like OpenTS are not updated as soon as the data is changed in real-time, do not show any history of the performance of different models and require that users themselves run the evaluations on models manually in order to update the leaderboard \citep{aksu2024gifteval}. 

\textbf{Development Overview} For this project, we focus on the development of a FRED-MD \citep{mccracken_fred-md_2016} pipeline to serve as the foundation of a continual source of data to be used on a recurring basis for updated testing.

To provide a standardized input interface for models to integrate into the leaderboard platform, we developed a schema for each script meant to test a model's forecasting. In addition to basic metadata regarding model name, type, etc, it specifies the appropriate dimensions of data frame for input into training/testing the model and the appropriate dimensions of the forecast output. The purpose of this standardization is to provide a unified interface for model testing. To store the model evaluations, we utilized Supabase, an open source platform. Model evaluations, once generated, are uploaded, and stored for aggregation and comparison, which is displayed via the frontend.

We utilized Next.js and Tailwind CSS to construct a user accessible leaderboard that allows for comparison across models in a straightforward manner, incorporating all of the evaluation needs that we require. In particular, we were able to implement rolling window analysis. This in sum allows for visualizing relative performance of models compared to each other, and to fully encapsulate the performance metrics we had previously set out.

\textbf{MVP Implementation} For our MVP, we implemented evaluations for forecasting on different horizons (F={12,24,36,48,60} months), a rolling forecasting strategy, and a look-back window of 96 months. For our non-foundation models, we make API calls to Nixtla’s MLForecast ~\citep{noauthor_mlforecast_nodate} and StatsForecast ~\citep{garza2022statsforecast} libraries to train and forecast on our dataset.

We created a fully functional, MVP version of the dynamic leaderboard by implementing the frontend and the aforementioned steps, allowing comparison across models with different time horizons, and rolling forecast evaluations. We compare an initial set of 8 models: historical average, LightGBM ~\citep{ke_lightgbm_2017}, ETS ~\citep{hyndman_chapter_2021}, TimesFM ~\citep{das_decoder-only_2024}, random forest ~\citep{breiman_random_2001}, nlinear ~\citep{zeng_are_2023}, linear, and linear regression approaches. 

\textbf{Next Steps} Based on a quick glance of our preliminary data, we have demonstrated variation across contexts (variation across time horizons for the models of interest and interesting differentiation for the various time horizon intervals) and seemingly significant trends in the relative performance of models over time in history, but we will have to do more formal statistical analysis to prove this significance.

With this framework, our next step is to swap in more interesting datasets in addition to FRED-MD. For example, FRED-QD focuses on quarterly macroeconomic data (e.g., GDP, consumption, investment) ~\citep{mccracken_fred-qd_2021}, and FRED-SD explores smaller subcategories like industrial production and housing ~\citep{bokun_fred-sd_2023}. 
Beyond the FRED family of datasets, additional sources like polling markets or prediction markets, can be valuable and are an untapped source of insight . This would necessitate the development of additional scripts for scraping new data sources.
Specific examples include Kalshi Prediction Markets ~\citep{beckhardt_harnessing_nodate} and ALFRED, which tracks revisions to past data points (e.g., updated GDP figures) ~\citep{stierholz_alfred_nodate}.

\section{AI Legislator: Eliciting Human Preferences for Creation of AI-Generated Policies}
\label{sec:ai_legislator}
The \textbf{AI Legislator} builds on the foundations laid by the \textbf{Economic Transformer}, integrating insights from multimodal time-series modeling and textual data analysis to understand the impacts of policy, establishing a framework for data-driven policymaking. The AI Legislator ensures that policy generation is grounded in both quantitative economic forecasting and qualitative stakeholder values.
By simulating diverse perspectives, it seeks to bridge ideological divides and enhance effective governance.

\subsection{Value Elicitation Framework for AI Policymaking: A Simulation-Based Approach}

This project will develop a framework for eliciting human values at scale to inform AI-driven policymaking. It builds on \cite{park2024generative} simulation of 1,052 individuals, achieving 85\% accuracy in replicating their survey responses, and uses this simulated environment to refine methods for capturing human moral and political preferences. The framework applies hierarchical Bayesian modeling, Moral Foundations Theory \citep{graham2013moral}, and optimal transport methods to represent and aggregate diverse values. By producing structured, empirically validated representations of values, the system can guide automated policy generators toward solutions that align with a broad range of stakeholder preferences.


\textbf{Objectives and Expected Significance}

Effective AI policymaking requires accurately representing and integrating human values into decision-making.

\textbf{Key challenges:}
\begin{itemize}
    \item Capturing value heterogeneity across diverse populations at scale
    \item Reconciling conflicting moral frameworks to produce actionable policy guidance
    \item Translating elicited values into formal constraints that inform automated policy generation
\end{itemize}


\textbf{Background and Technical Need}

Traditional methods for eliciting values, such as surveys or interviews, often face biases and cognitive overload, leading to inconsistent aggregate profiles. \cite{park2024generative} addressed this by using generative agents to simulate human attitudes and behaviors with measurable fidelity, enabling systematic refinement of elicitation techniques.

Our approach leverages the following key methodologies:

\textit{Strategic Query Design:} Aggregates value distributions into collective representations adaptively based on prior answers. \textit{Hierarchical Bayesian Modeling:} Models values at global, domain-specific, and individual levels, refining accuracy with fewer queries. \textit{Moral Foundations Theory:} Six core dimensions (Care/Harm, Fairness/Cheating, Loyalty/Betrayal, Authority/Subversion, Sanctity/Degradation, and Liberty/Oppression) provide a structured basis for queries and interpretation.


Strategic Query Design:

The system uses active learning to select queries that maximize information gain about users' moral foundations. Instead of generic questions, it employs targeted modalities such as binary policy choices (e.g., “Do you prefer stringent AI oversight or a permissive innovation environment?”),  resource allocation tasks (e.g., “Distribute 100 points among economic growth, equity, and AI safety”),  and moral dilemmas or scales (e.g., “Should AI prioritize individual freedom over collective welfare? Likert 1–7 scale”).

These queries, linked to moral dimensions like Liberty/Oppression or Fairness/Cheating, enable efficient updates of user-specific value parameters. Over multiple iterations, the hierarchical Bayesian model converges on a stable moral profile for each user.

Hierarchical Bayesian Modeling:
The model operates across three layers:
\begin{itemize}
    \item \textbf{Global Layer:} Captures universal moral dimensions as priors informed by Moral Foundations Theory \citep{graham2013moral}.
    \item \textbf{Domain Layer:} Adjusts these priors for specific policy contexts (e.g., data privacy in AI regulation).
    \item \textbf{Individual Layer:} Personalizes the distribution to the individual respondent. If responses indicate strong emphasis on Fairness, the model shifts that user’s distribution accordingly.
\end{itemize}

Using a posterior update rule, the model refines its estimates with each query, improving accuracy and efficiency:
\[
P(\theta_{\text{user}} \mid r_q) = \frac{P(r_q \mid \theta_{\text{user}}) P(\theta_{\text{user}})}{P(r_q)},
\]
where \(\theta_{\text{user}}\) represents the user’s value parameters and \(r_q\) is the observed response. By applying this rule iteratively, the system refines its estimates of a user’s moral foundations with each query.

Moral Foundations Contextualization:
Moral Foundations Theory \citep{graham2013moral} provides a structured basis for categorizing and interpreting responses across six core dimensions. This grounding ensures psychological validity and enhances clarity and reliability in inferred moral constructs, e.g., linking privacy vs. security debates to Authority/Subversion and Liberty/Oppression dimensions.

\textbf{Integration with Policy Generator}

The elicited values feed directly into the AI policy generator, serving as constraints or weights to align policies with identified moral priorities. For example, value profiles favoring fairness lead to policies promoting equity, while preferences for innovation may result in looser regulations with safety measures.

By integrating value elicitation into policymaking, the system identifies conflicts early and flags them for stakeholder deliberation. Simulations of demographic subgroup responses help policymakers anticipate potential unintended consequences.

\textbf{Timeline and Phases of Work}

\textit{1. Foundation Development:} Implement the hierarchical Bayesian model using Moral Foundations Theory as a baseline. Develop initial sets of queries corresponding to the six moral dimensions and create a simple prototype to demonstrate how value outputs guide a basic policy generator.

\textit{2. Simulation and Validation:} Deploy the elicitation framework on the agent population from \cite{park2024generative} to evaluate the accuracy and consistency of inferred values. Validate against known ground truths, adjust the query design and updating mechanisms, and conduct small-scale trials to verify stability of posterior distributions and reduction in uncertainty.

\textit{3. Integration and Scaling:} Finalize integration with the policy generator and optimize for real-world deployment to larger, more diverse populations. Pilot the framework in controlled settings to ensure practicality and reliability.

\textbf{Broader Implications}

This framework provides a systematic method for incorporating diverse moral perspectives into policymaking, enabling the creation of more inclusive policies. Its grounding in Moral Foundations Theory improves interpretability, helping policymakers understand the moral rationale behind policy outcomes. Future research could expand this approach to other domains, explore additional moral dimensions, or enhance statistical techniques for finer-grained value elicitation.

\subsection{Policy Generation for AI Legislator} 
This project aims to build an LLM-based policy generator that assists policymakers in addressing complex issues using simulated human perspectives for evaluation.  It will take as input a policy issue or query along with a distribution over values from the Value Elicitation project, and produce actionable, contextually-informed policy recommendations by leveraging inten decomposition, knowledge retrieval, simulated personas, and advanced language models. 

\textbf{Policy Generation Workflow.}

\textbf{1. Policy Issue and Intent Decomposition:} Policy issues retrieved from the frontend are translated into actionable goals via Intent Decomposition. A High-level policy intent \( H \) is parsed into actionable steps \( A = \text{Decompose}(H, \text{Context}) \), where the context includes relevant data \( D \) retrieved from the Knowledge Base \citep{dzeparoska2023}.

\textbf{2. Knowledge Base and Context Retriever:} The Context Retriever sources relevant information from static and dynamic Knowledge Bases. This contextual injection ensures the Core LLM operates with domain-relevant, and up-to-date information.

\textbf{3. Core LLM \& Policy Generation:} Using pretrained models (4o-mini, 4o) prompt engineering, and in later phases Retrieval-Augmented Generation (RAG), the LLM integrates contextual data to draft policies, where the policy \( P \) is generated as \( P = \text{CoreLLM}(Q, D) \), with \( Q \) representing the query and \( D \) the retrieved context.

\textbf{4. Policy Validation and Refinement:} The policy \( P \) is then assessed for feasibility, realism, and bipartisan appeal. Specifically, there should be three phases here:

\textit{Phase 1: Persona Querying} assesses demographic support \( S_p(P) \) and legislative feasibility \( S_l(P) \) through predefined personas, where the total score is calculated as \( S_{\text{total}}(P) = \alpha S_p(P) + \beta S_l(P) \). \textit{Phase 2: Transition to Agentic Framework} later in the development Representative and Mediator Agents are introduced for deeper refinement \citep{wang2023survey}, updating scores as \( S(P, G_i) \rightarrow S(P, G_i) + \Delta S(P, G_i) \).

\textbf{5. Output Formatting and Delivery:} Finalized policies are formatted into human-readable and API-ready outputs via the Policy Output Formatter and delivered to users through the frontend.

\textbf{6. Fine-Tuning:} Offline and periodic instruction tuning aligns models with policy goals, addressing gaps via data augmentation methods such as SMOTE and NLP-AUG \citep{chawla2002smote}.

\textbf{Broader Impacts.}  
This sub-project fosters civic engagement by integrating diverse perspectives into policy decisions, building trust in governance, and promoting economic welfare. Its methodologies for demographic simulation and feedback refinement align with Humanity Unleashed's mission to advance collaborative, AI-driven governance.

\section{Infrastructure for an AI-Driven Policy Interface}
\label{sec:policy_interface}

Alongside model development to support policymaker decision-making, we aim to create a platform that enhances user accessibility to U.S. legislative data (Section~\ref{subsec:frontend}). Interactions on this platform will then be integrated with the AI Legislator and Economics Transformer, enabling AI-driven policy generation and allowing users to have access to these inference capabilities (Section~\ref{subsec:backend}).

\subsection{Frontend Interface}
\label{subsec:frontend}

\textbf{Project Summary:} Our methodology combines the principles of Human-Computer Interaction (HCI) and advanced AI techniques like natural language processing (NLP) to create a user-centric platform that enhances the accessibility of U.S. legislative data. We will take a comprehensive, phased approach consisting of a user-centric design analysis, HCI integrations with a prototype, iterative usability testing, usability evaluation, and deployment. This will guide the development and iterative refinement of our platform, ensuring it effectively meets user expectations, reduces barriers to civic engagement, and access to legislative information.

\textbf{Objectives and Expected Significance:} This project addresses two crucial questions in civic technology: 
\begin{enumerate}
    \item How can we democratize complex legislative information, making it easily accessible to the public? 
    \item How can we leverage emerging generative AI capabilities to enhance and empower the democratic process instead of weakening it?
\end{enumerate}

\textbf{Background and Technical Needs:} Current platforms like congress.gov  \citet{congressgov} and govtrack.us  \citet{govtrack} offer valuable legislative data but the interfaces are overwhelming, hindering public engagement. The main challenges include poor user experience, limited AI-driven text simplification, and a lack of user-centered design. While past efforts have focused on backend data management, our approach applies recent HCI insights to redesign the presentation of legislative content. By leveraging NLP models, we aim to simplify bill language, making it clear and accessible, thereby filling a critical gap in civic engagement. \cite{legal_lang_digi_env}

\textbf{Research Description:} Our project aims to modernize legislative platforms like congress.gov and govtrack.us by combining user-centered design principles, Human-Computer Interaction (HCI) methodologies, and AI-driven tools. The result will be a user-friendly, accessible interface that democratizes access to legislative information, simplifies complex language, and fosters civic engagement.

\textbf{Phase 1: User-Centered Design Analysis:} Users that we envision finding this platform very useful may include policy think tanks, public policy students and engaged citizens of the United States. To ensure our platform addresses real user needs, we will begin with a thorough user-centered design analysis in collaboration with the product architecture team:
    
\textbf{Phase 2: HCI Integration and AI-Driven Text Simplification:} Using insights from Phase 1, we will design an intuitive interface with Figma, following HCI principles and the GOMS model \cite{three_paradigms_hci} to optimize user interactions. Key features likely include:

\begin{itemize}
    \item \textbf{Value Elicitation Interface:} Helps users understand where they lie on the political spectrum by answering a series of political questions.
    \item \textbf{Policy Generator: }Allows users to input topics or upload files for AI-generated policy drafts.
    \item \textbf{Toolbar and Modular UI: }Offers quick access to bill summaries, definitions, and translation tools. \cite{enhancing_ux}
    \item \textbf{Conversational Agent: }Simplifies legislative language in real time, appealing to younger audiences. \cite{web_design_diff_gen} \cite{survey_ui_prog}
    \item \textbf{NLP Simplification: }Provides plain-language explanations of legal texts using advanced transformer models. \cite{nlp_legal_tech}
\end{itemize}
    
\textbf{Phase 3: Iterative Testing:} Wireframes and prototypes will be refined through user feedback, focusing on usability metrics like task completion time and user satisfaction to address design flaws. \cite{func_flexibility}

\textbf{Phase 4: Usability Evaluation:} We will employ the \textbf{System Usability Scale (SUS),} a ten-question survey measuring ease of use and satisfaction. \cite{brooke_1996} We will also use \textbf{NASA TLX} to conduct interviews to assess cognitive workload and gather data for persona creation. \cite{nasa_tlx}

\textbf{Phase 5: Final Testing and Deployment:} Large-scale testing will be conducted through a human data campaign using a platform such as Prolific or Amazon Mechanical Turk. This approach will assess accessibility, performance, and user engagement, ensuring that the platform effectively meets diverse needs. We will use models like feedback loops will refine the design, setting a new benchmark for transparent and accessible civic tech interfaces. This approach integrates user-centered design, HCI insights, and AI-driven tools to create a transformative platform that simplifies legislative information and enhances public engagement.

\begin{figure}[!htb]
   \begin{minipage}{0.48\textwidth}
     \centering
     \includegraphics[width=1\linewidth]{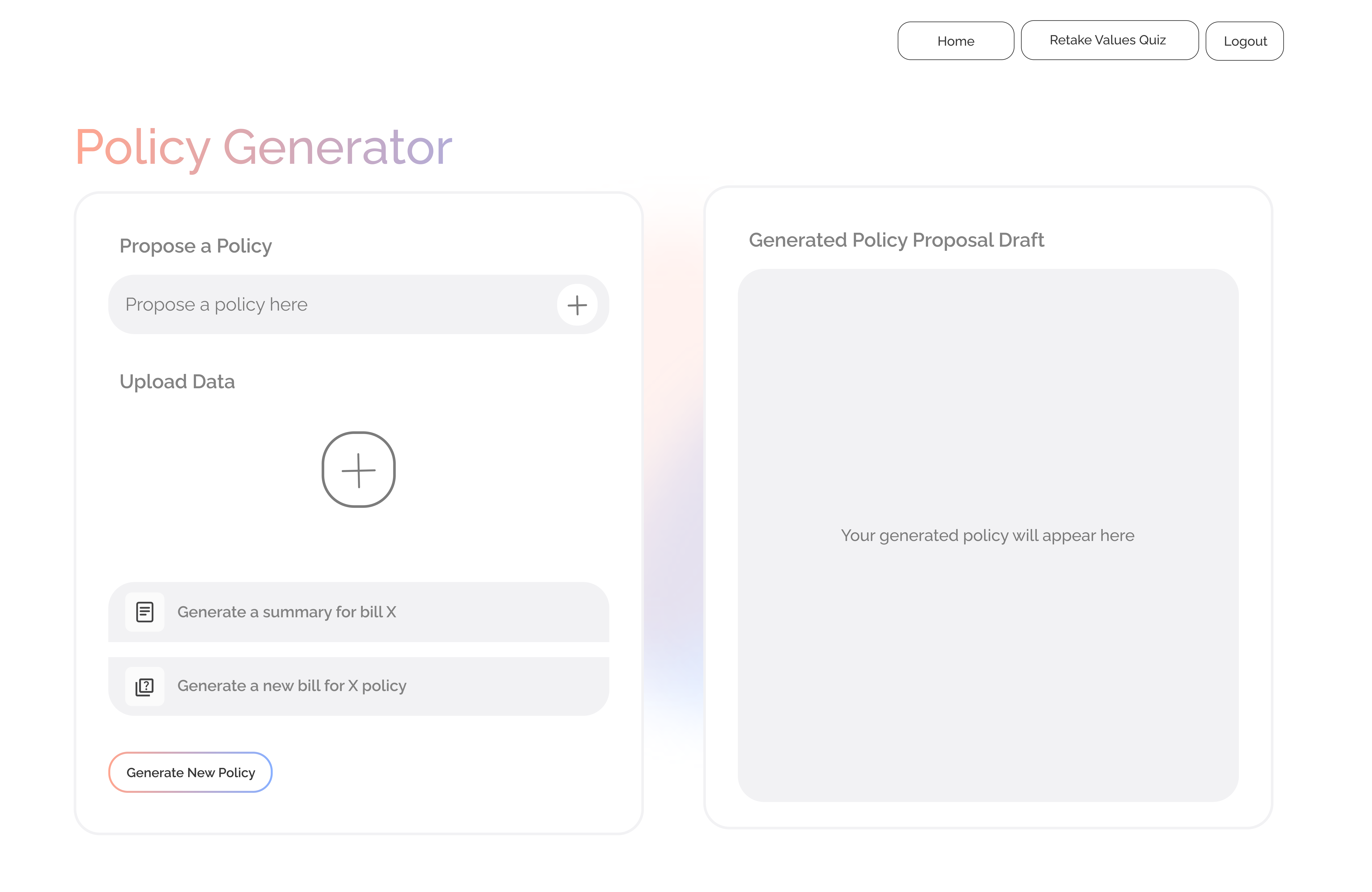}
     \caption{Policy Generator Model Frontend Mock-Up.}
   \end{minipage}\hfill
   \begin{minipage}{0.48\textwidth}
     \centering
     \includegraphics[width=1\linewidth]{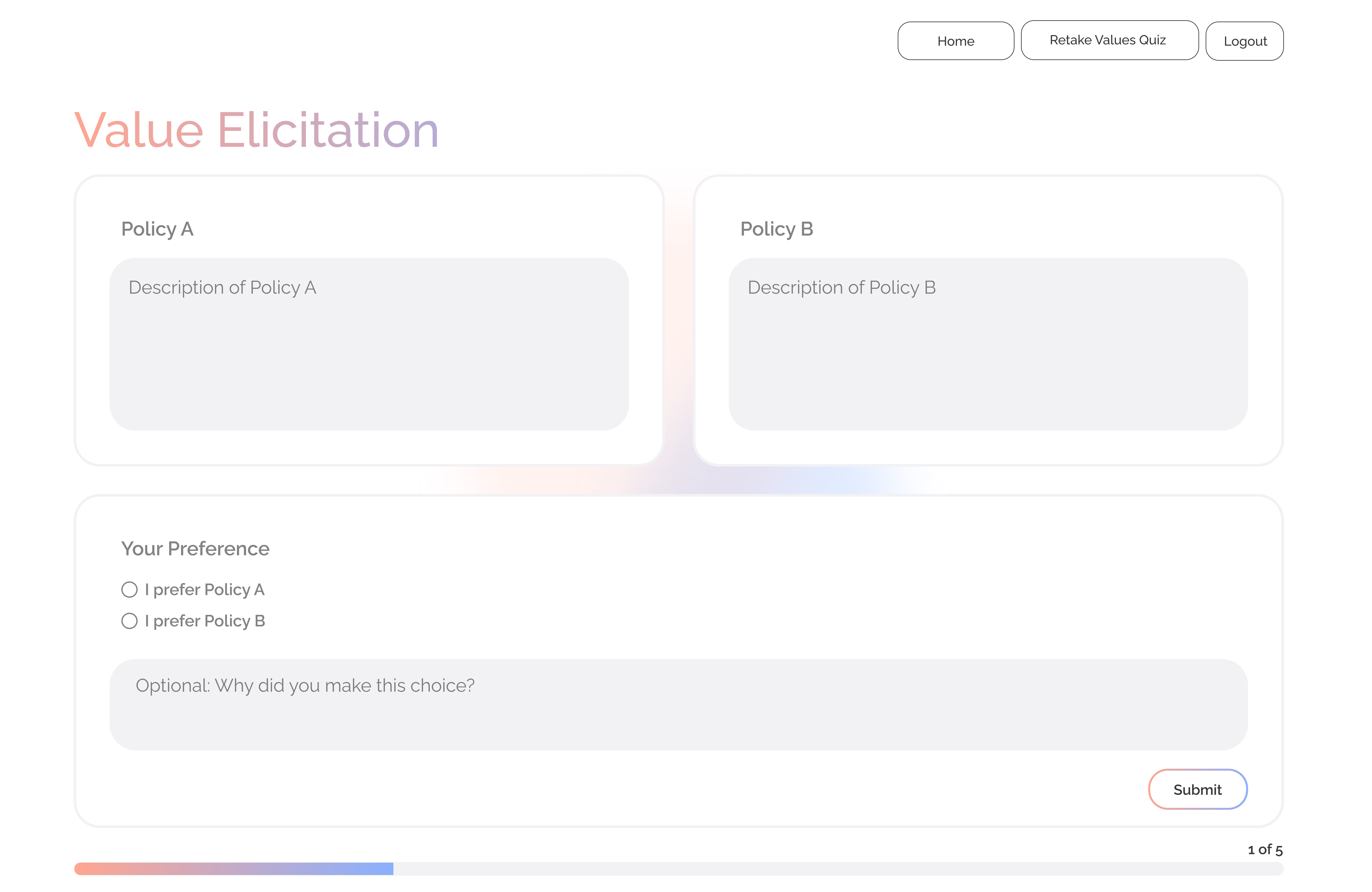}
     \caption{Value Elicitation Model Frontend Mock-Up.}
   \end{minipage}
\end{figure}

\textbf{Broader Impacts of the Proposed Work:} The outcomes of our work have the potential to shape future civic tech design standards, advocating for simpler and more inclusive interfaces for public information. The system could be adapted internationally, providing accessible legislative data across diverse political contexts and fostering global democratic participation. Additionally, our research on NLP simplification models has the promise to enhance educational technologies, bridging comprehension gaps in complex areas such as law and policy.

\subsection{Backend Implementation}
\label{subsec:backend}
This project aims to develop a robust and scalable backend system that seamlessly integrates the Policy Frontend, AI Legislator and Economics Transformer components within the Humanity Unleashed platform. This integration aims to facilitate AI-driven policy generation by enabling efficient data exchange, processing, and real-time interactions among system components. Furthermore, we will collect and store user-provided demographics information, policy preferences, and policy proposals as training data for the AI Legislator.


\textbf{Related Works:} Related works to this project include:
- govtrack.us, an open-source platform which tracks the activities of the US Congress, including bills and votes, legislators, and policies pertaining to personalized issues  (\cite{govtrack}).
- api.congress.gov, the official federal tool for obtaining bills in congress, their status, and legislator information (\cite{congressgov_api}).
- isidewith.com, a platform used by more than 80 million politically engaged citizens, which allows people to weigh in on political issues and show them which politicians share the most similar views (\cite{isidewith}). 

\textbf{API Design and Implementation:}  
The primary task is to develop a standardized API enabling actions such as authentication, database queries, and model inference. A consistent API schema ensures interoperability across components like the Economics Transformer and AI Legislator, which may have distinct input/output formats and performance profiles \cite{hmue2024microservices}. Prior work in microservices architecture has shown that well-defined interfaces and asynchronous communication patterns can simplify integration and maintenance at scale. Such approaches reduce coupling and facilitate incremental upgrades without disrupting the entire system \cite{nugroho2022clean}. The backend must process large volumes of policy data and support concurrent user interactions, necessitating efficient data handling, indexing, caching, and load balancing. As demonstrated by research in database design for microservices, NoSQL databases are effective in handling unstructured or semi-structured data at scale, ensuring rapid response times and robustness under heavy load \cite{soni2023sqlvsnosql}.

\textbf{Database Implementation:} The second critical responsibility is to implement a database to store user-related data. Optimizing the data pipeline is another challenge, as it must efficiently handle data ingestion, processing, storage, and retrieval. This includes managing multimodal inputs such as text and documents, both structured and unstructured. The backend should provide mechanisms for data validation, transformation, and normalization to ensure consistency and accuracy. Data handling for multimodal inputs necessitates flexible storage solutions. NoSQL databases like MongoDB are preferred for their ability to handle unstructured data (\cite{soni2023sqlvsnosql}).

\textbf{User and Data Security:} Security and transparency are also critical considerations. The backend must implement robust authentication and authorization mechanisms to protect sensitive policy data and user information. Ensuring data integrity and compliance with relevant regulations in our areas of operation is essential (\cite{yu2016bigprivacy}). Transparency in data processing and decision-making processes of AI components must be maintained to build user trust. Additionally, users must provide clear consent for the usage of their data in our models and be allowed to remove their data from our platform, to the best of our ability.

\textbf{Technical Implementation}

The proposed backend architecture adopts a microservices approach, utilizing FastAPI as the core framework for API development due to its high performance, ease of use, and support for asynchronous operations \cite{nugroho2022clean}. An API gateway will be implemented to act as a single entry point for all client requests, routing them to the appropriate microservices. This gateway will handle tasks such as authentication, rate limiting, and request validation.

\textbf{API Gateway Design Using FastAPI:} The API gateway will be built using FastAPI, leveraging its asynchronous capabilities to handle a high volume of concurrent requests efficiently. It will expose RESTful endpoints adhering to the OpenAPI 3.0 specification, ensuring standardization and ease of integration with other components. The gateway will implement middleware for logging, error handling, and security checks.

\textbf{Database Schema for Policy Storage:} For data persistence, MongoDB will serve as the primary database due to its flexibility in handling unstructured and semi-structured data. The database schema will include collections for users, policies, economic data, legislative information, and logs. Policies will be stored with fields such as \verb|policy_id|, \verb|title|, \verb|content|, \verb|author_id|, \verb|timestamps|, and \verb|metadata|. Indexing strategies will be employed to optimize query performance on frequently accessed fields.

\textbf{Real-Time Data Processing Pipelines:} To handle real-time data processing, an event-driven architecture will be adopted. Message brokers like RabbitMQ or Apache Kafka will facilitate asynchronous communication between services. For example, when a new policy proposal is submitted, an event will be published to a queue, triggering the Economics Transformer and AI Legislator services to process the data. This decouples services and enhances scalability.

\textbf{Security Implementation}

\textbf{JWT-Based Authentication:} Authentication will be managed using JSON Web Tokens (JWT), providing stateless and secure user authentication. Upon successful login, the system will issue a JWT containing user identity and role claims. JWT tokens will be securely generated using strong encryption algorithms (e.g., HS256 or RS256) and will include expiration times and claims to prevent token reuse and tampering. Secure storage of secret keys and regular key rotation policies will be enforced.

\textbf{Role-Based Access Control:} Role-Based Access Control (RBAC) will be implemented to manage authorization, defining roles such as admin, legislator, analyst, and general user, each with specific permissions. Sensitive endpoints will enforce authorization checks based on user roles. Access control checks will be implemented at the API gateway and service levels to ensure only authorized users can access certain endpoints or perform specific actions.

\textbf{Data Encryption Standards:} Data in transit will be secured using HTTPS with TLS 1.2 or higher (\cite{diyora2024blockchain}). Sensitive data at rest in the database will be encrypted using field-level encryption or full-disk encryption where appropriate. Encryption keys will be managed securely using key management services (e.g., AWS KMS or Azure Key Vault).

\textbf{User Data Privacy:} There is no comprehensive federal law governing the handling of political data. Political parties can collect voter data without explicit consent, purchase commercial data to enhance voter profiles, store data indefinitely, and use data for micro-targeting without restrictions (\cite{colin2013bennett}).

Some notable state-level privacy laws are:

- California (CCPA) requires companies to inform about personal data collected, allow opting out of data sales, provide the right to delete information, and applies to companies processing California residents' data regardless of location
- Virginia (CDPA; Consumer Data Protection Act) classifies political opinions as 'sensitive data', requires explicit consent for processing, and is stricter than CCPA for handling political data. Includes data protection assessments, opt-out option after consent, clear privacy notices, and regular review of consent validity
- Colorado’s Privacy Act focuses on transparency, clear disclosure of data collection purposes, right to opt out of data processing, and includes political preferences in protected categories.
All 50 states have breach notification laws. (\cite{Seun2024PrivacyLaws})

Therefore, we will enforce clear opt-out mechanisms and transparent information to any user regarding our use of their data in long-term storage and/or training models. Users will have to provide their clear consent to allow us to train models using their data. Once the model is trained, however, we will not be able to exclude or remove the user's data from the model weights. We will allow deletion of the user's data and account from our database at any point in time. In the highly unlikely event of a data breach, all affected users will be notified immediately.


\textbf{Monitoring and Logging Infrastructure:} A centralized logging system will be established using tools like ELK Stack (Elasticsearch, Logstash, Kibana) or Grafana for monitoring. Monitoring tools will provide dashboards for real-time insights into system health, resource utilization, and anomaly detection. Logs will capture request details, errors, performance metrics, and security events. Logs will include user identities, timestamps, actions performed, and the outcome. This facilitates compliance monitoring and forensic analysis in case of security incidents \cite{nugroho2022clean}.

\subsection*{Representative API Endpoints}

We include representative, high-level API endpoints served by the API Gateway in \autoref{app:api_endpoints}. These endpoints, adhering to RESTful principles, facilitate interactions such as user authentication, policy management, value elicitation, and economic forecasting. Some endpoints may provide real-time updates via WebSockets for asynchronous event notifications.

\textbf{Research Roadmap and Timeline:}  
    
    \textbf{Oct–Nov 2024 (Sprints 1–2):}  
    Set up the API gateway, implement JWT-based authentication, user management endpoints, and initial MongoDB schemas. 
    
    \textbf{Dec 2024–Feb 2025 (Sprints 3–4):}  
    Introduce containerization (Docker) and orchestration (Kubernetes) for horizontal scaling. Implement event-driven pipelines, asynchronous messaging, and indexing strategies. Validate features via product research and refine endpoints for policy CRUD and value elicitation. Integrate basic CI/CD pipelines.
    
    \textbf{Mar–Apr 2025 (Sprints 5–6):}  
    Enforce RBAC, data encryption at rest, and improved input validation. Set up performance monitoring, logging, and load testing. Optimize queries for stable performance. Submit workshop paper integrating backend feedback from other sub-teams.
    
    \textbf{May 2025 (Sprint 7+):}  
    Final integration testing with the full platform. Document APIs thoroughly (OpenAPI specs), prepare operational runbooks, and ensure readiness for conference submissions and public release.
    

\section{Analyzing the Employment and Social Welfare Impacts of GPT and LLM Adoption }
\label{sec:societal_impact}


The societal impact of LLMs on the U.S. economy are critical to contextualizing this project. As informed by the integrated AI policymaking platform described previously, we analyze economic indicators, labor markets, and public trust, to identify key risks and opportunities arising from LLM adoption. This assessment guides strategies for minimizing harms and maximizing inclusive benefits through our proposed approach.

\textbf{Overview:} This proposal aims to investigate the impact of Generative Pre-trained Transformers (GPT) and Large Language Model (LLM) adoption on employment levels and social welfare within the United States economy. By reassessing existing research with updated models and data, we seek to understand how these advanced technologies affect labor markets. Furthermore, we will explore policy interventions using a Transferable Utility framework to mitigate adverse social consequences, ensuring equitable outcomes for displaced workers. In doing so, our analysis complements the broader mission of the larger initiative—our findings on LLM-driven economic shifts will inform the Economics Transformer’s forecasting capabilities and guide the AI Legislator’s policy recommendations. Further, by communicating these insights, we also aim to raise public awareness of the far-reaching societal impacts of advanced AI, fostering more informed discourse and engagement around the development of equitable, data-driven policy solutions.

\textbf{Introduction:}
The rapid advancement of artificial intelligence (AI) technologies, particularly GPT and LLMs, has transformative potential across various industries. These technologies are increasingly capable of performing tasks that were traditionally the domain of human workers, raising concerns about potential job displacement and its impact on social welfare \citep{FreyOsborne2017, BrynjolfssonMcAfee2014}. While automation has historically led to both job destruction and creation \citep{Autor2015}, the scale and speed of GPT and LLM adoption necessitate a thorough examination of their implications for the labor market.

This research seeks to analyze the effects of GPT and LLM technologies on employment levels and social welfare in the United States. Additionally, we aim to identify policy measures that can address the social consequences of workforce reductions, ensuring that the benefits of technological advancements are equitably distributed

\textbf{Primary Research Question:}
What is the impact of GPT and LLM adoption on employment levels and social welfare within the United States economy? How can policymakers address the social consequences of workforce reductions due to GPT technologies to ensure equitable outcomes? Our project aims to address this in parts:

    1. \textbf{Empirical Reassessment:} Building upon the current \textit{GPTs are GPTs} paper, we aim to reassess their question with updated models and data. Specifically, we will analyze how the adoption of GPT and LLM technologies impacts employment levels and social welfare within the United States economy. \\
    2. \textbf{Policy Design and Evaluation:} Utilizing a Transferable Utility (TU) framework, we will explore how policymakers can design taxation and social insurance policies to redistribute the excess utility gained by companies to workers adversely affected by GPT-induced job displacement. Specifically, we will examine:
    
        • The role of employer subsidies in incentivizing workforce retention. \\
        • How the proliferation of LLMs influences the Marginal Value of Public Funds (MVPF) associated with these interventions. \\
        • How shift-exposure changes across economic sectors.

\textbf{Literature Review:} The debate on automation and employment has been longstanding, with early concerns dating back to the Industrial Revolution. Recent studies have focused on the susceptibility of jobs to computerization \citep{FreyOsborne2017} and the historical perspective of workplace automation \citep{Autor2015}. \citet{BrynjolfssonMcAfee2014} discuss the implications of the "Second Machine Age," highlighting the potential for significant technological unemployment.

Agent-based computational economics provides a framework for modeling complex economic systems with heterogeneous agents \citep{Tesfatsion2006}. This approach can capture the micro-level interactions that give rise to macroeconomic phenomena, which is particularly relevant when assessing the impact of GPT and LLM technologies on labor markets.

Policy interventions have been proposed to mitigate the adverse effects of automation on employment. The Marginal Value of Public Funds (MVPF) concept, as discussed by \citet{HendrenSprung-Keyser2020}, offers a method for evaluating the cost-effectiveness of government policies aimed at improving social welfare. 

\textbf{Empirical Analysis and Modeling:} Our methodology integrates theoretical modeling with empirical analysis to provide a comprehensive understanding of the impact of GPT and LLM adoption.  Our empirical strategy involves:

    • \textbf{Data Collection:} Compiling data from O*NET on occupational characteristics, BLS for employment and wage statistics, IRS data for income and tax information, and annotations on task automation potential from both human experts and GPT models. \\
    • \textbf{Shift-Share Analysis:} Employing Bartik shift-share instruments to measure industry and regional exposure to GPT/LLM technologies, following the methodologies of \citet{GoldsmithPinkham2020} and \citet{Borusyak2022}. \\
    • \textbf{Econometric Modeling:} Estimating the causal impact of GPT/LLM adoption on employment and wages using instrumental variable regression techniques to address endogeneity concerns. \\
    • \textbf{Robustness Checks:} Performing sensitivity analyses to ensure the validity of our findings, including alternative specifications and placebo tests.

\textbf{Metrics:} To evaluate the impact and effectiveness of policy interventions, we will utilize the following metrics, and possibly more:

Bartik Shift-Share Exposure Variable serves as a critical tool for decomposing employment changes and analyzing the regional or industry-level effects of GPT/LLM adoption. This method involves calculating the extent to which specific regions or industries are exposed to GPT/LLM adoption based on their initial employment composition \citep{Adao2019}. Additionally, it enables the attribution of employment changes to either GPT/LLM adoption or other economic factors, ensuring that causal relationships are properly identified \citep{BorusyakHull2023}.

Kaldor-Hicks Efficiency is employed to assess whether the economic gains from GPT/LLM adoption can compensate for the losses incurred by displaced workers. This metric focuses on welfare evaluation by measuring the net change in social welfare, aggregating the gains and losses of all stakeholders. Furthermore, it explores compensation mechanisms that, in theory, could ensure no stakeholder is left worse off, thus adhering to the criteria for Kaldor-Hicks efficiency \citep{Mishan1981}.

Pareto Efficiency and $\epsilon$-Pareto Efficiency are integral for evaluating the efficiency of resource allocation in the context of GPT/LLM adoption. Pareto Efficiency identifies allocations where no individual can be made better off without making someone else worse off. In contrast, $\epsilon$-Pareto Efficiency allows for minimal efficiency losses to achieve greater equity, acknowledging the trade-offs between efficiency and equity \citep{Varian1974}.

Marginal Value of Public Funds (MVPF) quantifies the cost-effectiveness of government interventions aimed at mitigating the societal impacts of GPT/LLM adoption. MVPF is computed by estimating the social benefit per dollar of government expenditure, as outlined by \citet{HendrenSprung-Keyser2020}. This metric is particularly useful for comparing different policy interventions—such as subsidies, retraining programs, and social insurance—and prioritizing those that maximize social welfare.

\textbf{Data:} Our analysis draws on several critical data sources to ensure comprehensive evaluation and accurate modeling. The O*NET database provides detailed information on job requirements and worker attributes, enabling an in-depth analysis of task automation potential. Employment and wage data from the Bureau of Labor Statistics (BLS) offer insights into the distribution of occupations and industry-level dynamics. IRS data contribute income and tax information essential for evaluating fiscal policy impacts. Finally, human and GPT-annotated labels are employed to assess task susceptibility to automation by GPT/LLM technologies, enhancing the precision and reliability of our exposure measures.  

\section{Conclusion and Future Directions}
\label{sec:conclusion}

This research proposal outlines a comprehensive effort to build a cooperative AI policymaking platform that integrates advanced multimodal time-series forecasting with large language model (LLM)-driven value elicitation and policy generation. By coupling these emerging technical capabilities with a visually appealing frontend the proposed system aims to support more informed, data-driven decision-making for the public good.

\subsection*{Summary of Contributions}
Our approach centers on developing a multimodal ``Economics Transformer''---a specialized time-series LLM capable of assimilating numeric economic indicators, policy documents, and event data to forecast the impacts of proposed policies. We pair this with an ``AI Legislator'' component that elicits user values, learns nuanced public preferences, and proposes evidence-based policy interventions grounded in these values to feed to the ``Economics Transformer". Taken together, these advancements promise a platform that hopes to predict economic and social outcomes more accurately along with actively incorporating public input, enhancing legitimacy and inclusiveness in policymaking.

\subsection*{Expected Outcomes and Impact}

We primarily aim to achieve the following outcomes:

    \textbf{Technical Benchmarks and Open-Source Tools:} We expect to deliver baseline models, code repositories, and datasets as reference points for researchers in AI, economics, and public policy. By releasing these resources openly, we aim to foster a community of contributors who can refine and extend our methods.
    
    \textbf{Improved Policy Forecasting Models:} The integrated model architecture---combining textual and temporal data---should yield richer predictive insights, surpassing traditional forecasting approaches in accuracy, uncertainty quantification, and robustness.
    
    \textbf{Inclusive Value Elicitation and Policy Proposals:} Through value elicitation and user modeling, we seek to produce policy recommendations that better reflect a diverse range of public values. Ensuring that recommendations consider broad stakeholder input can lead to more equitable outcomes and enhanced trust in AI-driven policy support tools.

\subsection*{Challenges and Mitigation Strategies}
Developing such a platform will not be without significant challenges. Data integration may require sophisticated preprocessing and careful alignment of temporal and textual sources. The computational cost of training large multimodal models is substantial, and we will rely on efficiency strategies---such as parameter-efficient fine-tuning and mixture-of-experts architectures---to manage complexity. Additionally, safeguarding fairness and mitigating bias in model outputs will be critical to maintaining broad support. We will employ rigorous auditing, transparency measures, and fairness constraints in both forecasting and policy generation modules.

\subsection*{Scalability and Generalization}
Our methodology is designed to scale across policy domains, from economic regulation to environmental policy, and from local governance to international negotiations. By developing extensible, open-source architectures and tooling, we encourage the adaptation of our models for diverse regional contexts, policy areas, and data regimes.

\subsection*{Long-Term Vision and Future Work}
This project lays the groundwork for next-generation AI policymaking support systems. Future directions include refining multimodal modeling techniques, exploring causal inference and counterfactual analysis for robust policy impact assessment, and expanding participatory mechanisms for public engagement. Further research may incorporate richer modalities---such as geospatial data or real-time sentiment streams---and integrate more advanced deliberation frameworks to facilitate constructive dialogue among policymakers, domain experts, and the public.

In sum, this proposal envisions a future where AI systems not only assist in forecasting the consequences of policy decisions but also guide the creation of more equitable, data-driven policies. By opening our code, methods, and datasets to the global research community, we aspire to catalyze a new era of transparent, accountable, and inclusive policymaking enhanced by cooperative artificial intelligence.

\clearpage

\bibliography{sample}

\clearpage

\appendix

\section{High-Level Backend API Endpoints}\label{app:api_endpoints}

\textbf{User \& Authentication:}
\begin{itemize}
    \item \texttt{POST /auth/login}: Returns a JWT upon successful credential verification.
    \item \texttt{POST /auth/logout}: Invalidates active tokens.
    \item \texttt{GET /users/me}: Retrieves the authenticated user’s profile and role information.
\end{itemize}

\textbf{Policy Management:}
\begin{itemize}
    \item \texttt{POST /policies}: Submit a new policy draft (requires \texttt{policy\_id}, \texttt{content}, \texttt{metadata}).
    \item \texttt{GET /policies/\{policy\_id\}}: Retrieve policy details, including economic forecasts and AI-generated recommendations.
    \item \texttt{PUT /policies/\{policy\_id\}}: Update policy draft content or metadata (restricted to authorized roles).
    \item \texttt{POST /policies/\{policy\_id\}/finalize}: Signal that a policy is ready for legislative evaluation and public dissemination (admin-only).
\end{itemize}

\textbf{User Value Elicitation:}
\begin{itemize}
    \item \texttt{POST /values/responses}: Submit user answers to a value elicitation questionnaire.
    \item \texttt{GET /values/preferences}: Retrieve aggregated or personalized preference data for a user or cohort.
\end{itemize}

\textbf{Economic Analysis \& Forecasting:}
\begin{itemize}
    \item \texttt{GET /economics/indicators}: List available economic indicators (e.g., GDP, inflation).
    \item \texttt{GET /economics/forecast?policy\_id=X}: Trigger or retrieve an economic forecast for the specified policy scenario.
\end{itemize}

\textbf{Legislative Data Integration:}
\begin{itemize}
    \item \texttt{GET /legislation/bills}: Query recent bills, amendments, or public laws.
    \item \texttt{GET /legislation/bills/\{bill\_id\}}: Retrieve detailed legislative text, metadata, and related amendments.
\end{itemize}

\textbf{Real-Time Updates \& Notifications:}
\begin{itemize}
    \item \texttt{GET /events/subscribe} (WebSocket): Subscribe to event-driven updates (e.g., completion of a forecast, changes in a policy’s status).
\end{itemize}

\end{document}